**Valley Isospin Controlled Fractional Quantum Hall States in Bilayer Graphene**

Ke Huang[1,2], Hailong Fu[1], Danielle Reifsnyder Hickey[3,4], Nasim Alem[4], Xi Lin[2,5,6], Kenji Watanabe[7], Takashi Taniguchi[8], Jun Zhu[1,9*]

**Affiliations**

[1]Department of Physics, The Pennsylvania State University, University Park, Pennsylvania 16802, USA.

[2]International Center for Quantum Materials, Peking University, Beijing 100871, China

[3] Department of Chemistry, The Pennsylvania State University, University Park, Pennsylvania 16802, USA

[4] Department of Materials Science and Engineering, The Pennsylvania State University, University Park, PA 16802, USA

[5]Beijing Academy of Quantum Information Sciences, Beijing 100193, China

[6]CAS Center for Excellence in Topological Quantum Computation, University of Chinese Academy of Sciences, Beijing 100190, China

[7]Research Center for Functional Materials, National Institute for Materials Science, 1-1 Namiki, Tsukuba 305-0044, Japan.

[8]International Center for Materials Nanoarchitectonics, National Institute for Materials Science, 1-1 Namiki, Tsukuba 305-0044, Japan.

[9]Center for 2-Dimensional and Layered Materials, The Pennsylvania State University, University Park, Pennsylvania 16802, USA.

[*]Correspondence to: jzhu@phys.psu.edu (J. Zhu)

A two-dimensional electron system placed in a magnetic field develops Landau levels, where strong Coulomb interactions lead to the appearance of many-body correlated ground states. Quantum numbers similar to the electron spin enable the understanding and control of complex ground state order and collective excitations. Owing to its spin, valley and orbital degrees of freedom, Bernal-stacked bilayer graphene offers a rich platform to pursue correlated phenomena in two dimensions. In this work, we fabricate dual-gated Bernal-stacked bilayer graphene devices and demonstrate unprecedented fine control over its valley isospin degrees of freedom using a perpendicular electric field. Higher sample quality enables us to probe regimes obscured by disorder in previous studies. We present evidence for a new even-denominator fractional quantum Hall state at filling factor $\nu$ = 5/2. The 5/2 state is found to be spontaneously valley polarized in the limit of vanishing valley Zeeman splitting, consistent with a theoretical prediction made regarding the spin polarization of the Moore-Read state. In the vicinity of the even-denominator fractional quantum Hall states, we observe the appearance of the predicted Levin-Halperin daughter states of the Moore-Read Pfaffian wave function at $\nu$ = 3/2, 7/2, and - 1/2 and of the anti-Pfaffian at $\nu = 5/2$. These observations suggest the breaking of particle-hole symmetry in bilayer graphene. We construct a comprehensive valley polarization phase diagram for the Jain sequence fractional states surrounding filling factor 3/2. These results are well explained by a two-component composite fermion model, further demonstrating the SU(2) nature of the valley isospin in bilayer graphene. Our experiment paves the path for future efforts of manipulating the valley isospin in bilayer graphene to engineer exotic topological orders and quantum information processes.

## I. INTRODUCTION

Electrons occupying a partially filled Landau level (LL) experience strong Coulomb interactions that lead to a plethora of correlated electronic states, generally known as the fractional quantum Hall (FQH) effect. The FQH effect hosts complex many-body wave functions, non-trivial topology, and unconventional quantum exchange statistics that are potentially useful for topological quantum computing[1-9]. In particular, even-denominator states occurring at half fillings, such as the $\nu = 5/2$ state in GaAs[1,2,6], have attracted ongoing attention of the field since the 5/2 state is postulated to be a *p*+i*p* superconductor and harbors non-Abelian excitations potentially useful in the construction of a topological qubit[3,5,10]. Its fundamental novelty and technological appeal have motivated many studies and the discoveries of even-denominator FQH states with potentially similar origins in other 2D systems such as bilayer graphene[11-13], ZnO[14], and $WSe_2$ [15]. While thermal conductance measurements have provided good evidence on the non-Abelian nature of the 5/2 state [16], its exchange statistics has not been explicitly verified in interferometry studies. The particle-hole symmetry of the even-denominator states is fundamental to its understanding as ground states of different symmetries belong to different topological orders and harbor different edge modes.[2,17-22] Furthermore, the Moore-Read wave function of the 5/2 state must be a spin triplet and is in fact expected to be spontaneously spin polarized in the limit of zero Zeeman splitting[3,10,23,24]. Experimental studies of these important issues remain ongoing [6].

The electron spin and spin-like degrees of freedom in new materials and artificial structures play a fundamental role in the formation of correlated phenomena through exchange interactions and the symmetry requirement of a many-body wave function[25-28]. Owing to its spin, orbital and valley isospin degrees of freedom, the $E = 0$ octet of bilayer graphene (BLG) supports a rich variety of spontaneously broken symmetries and correlations[11,12,29-31]. In particular, a perpendicular electric displacement field $D$ generated through dual gating controls the *electrical* valley Zeeman splitting $E_v$ between states occupying different valley/layers, thus offering a powerful experimental knob to control the characters of the LLs and the nature of the interactions they support. This tuning is compatible with device scaling and is independent of the magnetic field $B$, which controls the strength of the Coulomb interactions. It is a distinct experimental control of the BLG platform, which can be deployed to construct and elucidate many-body phenomena in a correlated, multi-component 2D system.

In this work, we have made ultra-high-quality BLG devices that enable fine control of the valley Zeeman splitting $E_v$ and explored its profound impact in realizing FQH states with different ground state orders and valley isospin (VIS) polarizations. In the regime of low D-field, an unprecedented even-denominator FQH state emerges at filling factor 5/2 and is found to be spontaneously valley polarized in the limit of vanishing $E_v$. We observe clear Levin-Halperin daughter states of either the Moore-Read Pfaffian or the anti-Pfaffian in the vicinity of $\nu$ = 7/2, 3/2, -1/2 (Pfaffian), and $\nu$ = 5/2 (anti-Pfaffian), with the appearance of a Hall resistance plateau at $\nu$ = 1+7/13. These results suggest that the even-denominator FQH states in BLG break particle-hole symmetry and the broken symmetry sensitively depends on the underlying interactions. We construct a comprehensive experimental phase diagram of the VIS polarization for odd-denominator FQH states between $0 < \nu < 2$. These measurements truly establish the valley isospin in BLG as a spin-like electronic degree of freedom and pave the pathway to future efforts exploiting its utility in controlling the ground state order and topology of correlated electronic states.

## II. DEVICE FABRICATION

Our dual-graphite-gated, h-BN encapsulated Hall bar devices are made using dry van der Waals transfer and side contact techniques largely following methods introduced in the literature[11,12,32]. A different etching protocol is used in our fabrication process, which led to higher-quality electrical contacts compared to prior studies[11]. The details of the fabrication are given in Appendix A. Figure 1(a) shows an optical micrograph of device 002. The carrier density $n$ and displacement field $D$ are respectively given by $n = g_{\mathrm{BG}}(V_{\mathrm{BG}} - V_{\mathrm{BG0}}) + g_{\mathrm{TG}}(V_{\mathrm{TG}} - V_{\mathrm{TG0}})$, and $D = [g_{\mathrm{BG}}(V_{\mathrm{BG}} - V_{\mathrm{BG0}}) - g_{\mathrm{TG}}(V_{\mathrm{TG}} - V_{\mathrm{TG0}})]e/2\varepsilon_0$, with the gating efficiencies $g_{\mathrm{BG}}$ = $7.3\times10^{11}$ V$^{-1}$ cm$^{-2}$ and $g_{\mathrm{TG}}$ = $5.9\times10^{11}$ V$^{-1}$ cm$^{-2}$ in device 002. Measurements used to characterize the devices, as well as parameters of devices 011 and 015 are also given in Appendix A.

## III. RESULTS AND DISCUSSION

### A. An even-denominator fractional quantum Hall state at $\nu$ = 5/2

Figure 1(c) shows a false color map of the longitudinal resistance $R_{xx}$ ($D$, $\nu$) in the filling factor range $1 < \nu < 4$ at $B$ = 18 T. Integer and fractional quantum Hall states appear as dark lines. They occupy the $|\pm0\rangle$ and $|\pm1\rangle$ LLs of the BLG, the wavefunction of which are illustrated in Fig. 1(b)[29]. Figure 1(d) gives an energy level diagram of this regime. The valley Zeeman energy $E_{\mathrm{v}}^{\pm}$ is defined as $E_{\mathrm{v}}^{\pm} = \pm1/2\ g_v D$, where $\pm1/2$ correspond to the quantum numbers of the two valley isospins and $g_v$ is the bare valley $g$-factor. The valley Zeeman splitting $E_v$ increases with increasing $D$ following $E_v = g_v D$. We obtain the value of $g_v$ = 1.43 K/(mV/nm) from Fig. 8 in Appendix B and Ref [30]). The $|+0\rangle$ and $|-1\rangle$ levels become degenerate at $D = D^*$, where $E_v$ = $E_{10}$[11,12,29,30]. This level crossing leads to the closing of gaps for states occupying these two LLs, resulting in an $R_{xx}$ increase in our measurements. We mark the $D^*$ transitions in Fig. 1(c) using four red dashed lines. In regimes of $D > D^*$, we observe the same LL orbitals as in previous studies, where two even-denominator FQH states at $\nu$ = 3/2 and 7/2 have been identified[11,12]. They also occur in our devices (See Fig. 1(c)). In this work, we focus on the regime of $D < D^*$, where disorder has obscured previous studies[11,12].

In this small-$D$ regime, states in the range of $2 < \nu < 4$ occupy the $|\pm1\rangle$ LL levels. Remarkably, a strong $R_{xx}$ minimum develops at $\nu$ = 5/2 in our devices. This $R_{xx}$ minimum occurs in multiple devices and is accompanied by well-quantized $R_{xy}$ plateau as shown in Fig. 9(a) of Appendix B. Thus, the state at $\nu$ = 5/2 is an even-denominator FQH state. Its appearance on the $|\pm1\rangle$ LL levels, while somewhat intuitive, is only observed now thanks to the high quality of our devices. As the LL diagram in Fig. 1(d) shows, the 5/2 state resembles its counterpart in GaAs, but with the spin index now replaced by the VIS index in BLG. A Moore-Read wave function requires the 5/2 state to be polarized in any spin and isospin sectors while numerical simulations of a zero-thickness 2D system further support its spontaneous polarization in the limit of $E_z = 0$ [10,23,24]. Numerous experiments have examined the spin polarization of the 5/2 state in GaAs [6,33-40] . While the state is generally thought to be spin polarized in a sufficiently large magnetic field, measurements conducted in the low-field range of $B \leq 1$ T have uncovered potential phase transitions[36,37]. In the small-$E_z$ regime, the interpretation of measurements is complicated by the finite wave function thickness in GaAs and the changes of the interaction energies when the magnetic field changes. In comparison, the electrical tuning of $E_v$, together with the high quality and near-zero thickness of BLG presents a clean approach to probe the valley isospin polarization in BLG, in advance of theoretical calculations.

We measure the gap of the 5/2 state $\Delta_{5/2}$ as a function of the $D$-field using thermally activated transport. Figure 2(a) plots the $D$-field sweeps of $R_{\mathrm{xx}}^{5/2}(\widetilde{D})$ measured at different temperatures, from which we obtain data points for $R_{xx}(T)$ and plot them in an Arrhenius plot in Fig. 2(b). Fits to $R_{xx}(T) \sim \exp(-\Delta/2k_BT)$ yield the gaps $\Delta_{5/2}(\widetilde{D})$ at different $D$-fields. Figure 2(c) plots $\Delta_{5/2}(\widetilde{D})$ for both positive and negative values in the range of $|\widetilde{D}| > 1 - 2$ mV/nm, where the effect of potential disorder can be neglected. Here $\widetilde{D} = D + 12.5$ mV/nm adjusts for a small offset in the applied $D$-field. The bottom x-axis of Fig. 2(c) plots the normalized $E_v/E_c$, where $E_c = \frac{e^2}{4\pi\epsilon\epsilon_0 l_B} = 217\sqrt{B[\mathrm{T}]}$ K is the Coulomb energy scale in BLG using $\epsilon = 3$ for the dielectric constant of h-BN. The gap $\Delta_{5/2}(\widetilde{D})$ saturates at 0.3/0.35 K respectively at large positive/negative $D$-field and extrapolates to a finite value of approximately 0.2 K at $\widetilde{D} = 0$ from both sides. The trend $\Delta_{5/2}(\widetilde{D})$ exhibits, together with a sizable gap at $E_v = 0$, provides strong experimental evidence for a spontaneously valley polarized 5/2 state. As we will demonstrate in Figure 4, the valley isospin in BLG indeed behaves as an SU(2) spin. The spontaneous valley polarization of the 5/2 state is consistent with numerical simulations performed for the real electron spin of the Moore-Read State [23,24]. Prior work in BLG finds the even-denominator state at $\nu$ =3/2 to be spin polarized at 17 T [11]. Given the similar Zeeman splitting in both cases, we expect the 5/2 state examined here to be also spin polarized, thus satisfying the triplet requirement of the Moore-Read wave function in any spin or isospin [3].

Figure 2(d) shows the behavior of $R_{\mathrm{xx}}^{5/2}(\widetilde{D})$ and $\Delta_{5/2}(\widetilde{D})$ close to $\widetilde{D}=0$, where the valley polarization of the 5/2 state is expected to switch abruptly in samples free of disorder. Instead, we observe a percolation transition from the $|+1\rangle$ LL to the $|-1\rangle$ LL with a full-width of approximately $\delta D \sim 0.7$ mV/nm or $\delta E_{\mathrm{v}} \sim 1$ K. This transition is signaled by two approximately $T$-independent crossover points at $D_{\mathrm{c}} \sim \pm 0.35$ mV/nm in Fig. 2(d), where $\Delta_{5/2}$ drops to zero apparently. We emphasize that the $T$-dependence of $R_{\mathrm{xx}}$ close to $\widetilde{D}=0$ is not related to the physics of the 5/2 state but rather reflects the general conduction behavior of a percolation network; The increased bulk conduction in the presence of mixed valley domains leads to $R_{\mathrm{xx}} \sim \sigma_{\mathrm{xx}}^{-1}$, hence, an opposite d$R$/d$T$ near $\widetilde{D}=0$. The low disorder of our devices allows us to probe the intrinsic behavior of the 5/2 state down to the small valley Zeeman splitting of $E_{\mathrm{v}}/E_{\mathrm{c}} \sim 10^{-3}$, where the 5/2 state remains valley polarized.

**B. The particle-hole symmetry of the even-denominator fractional quantum Hall states**

The particle-hole (p-h) symmetry of a half-filled LL is another open question of keen interest to the quantum Hall community [2,16-22,41-45]. The Moore-Read Pfaffian and its p-h conjugate the anti-Pfaffian explicitly break the p-h symmetry. Their energy difference is small and each state has gained support in numerical calculations performed in GaAs [17,18,41-43]. Theory also suggests the possibility of a third ground state, known as the p-h Pfaffian, which preserves the p-h symmetry and can be stabilized when both disorder and LL mixing are included [44,45]. The different Majorana neutral modes of the three distinct topological orders allow them to be differentiated in thermal conductance and noise measurements and recent experimental results have favored the p-h Pfaffian [16,22]. However, the situation is far from being settled. In the hierarchical theory, quasiparticles and quasiholes away from the half-filled N = 1 LL condense into incompressible fractional quantum Hall states known as the Levin-Halperin states [46]. The first daughter states of the Pfaffian occur at partial fillings $\tilde{\nu}$ = 7/13 and 8/17 while the first daughter states of the anti-Pfaffian occur at $\tilde{\nu}$ = 6/13 and 9/17 [46]. Transport studies in GaAs have revealed the appearance of a clear FQH state at $\tilde{\nu}$ = 6/13, suggesting an anti-Pfaffian ground state [21]. The understanding and reconciling of these measurements remain ongoing. BLG exhibits multiple even-denominator FQH states, including the new 5/2 state [11,12]. Using the Levin-Halperin states as indicators, our transport studies point to the Pfaffian order at $\nu$ = 3/2, 7/2, -1/2 and the anti-Pfaffian order at $\nu$ = 5/2.

The p-h asymmetry appears to be most prominent at $\nu$ = 3/2, as shown in Fig. 3(a) for device 015 and Fig. 11(a) for device 002. We see clear $R_{\mathrm{xx}}$ minima at $\tilde{\nu}$ = 7/13 and 8/17, together with a Hall plateau at $R_{\mathrm{xy}}$ =0.65 $h/e^2$ corresponding to the 20/13 filling. Strong $R_{\mathrm{xx}}$ minima at these fractional fillings are also found in device 002 and shown in Fig. 11(a). In the literature, a peak at 7/13 was also found in capacitance measurements of Ref.[12]. In contrast, the newly observed 5/2 state appears to follow the anti-Pfaffian order, as suggested by the appearance of $R_{\mathrm{xx}}$ minima at $\tilde{\nu}$ = 6/13 and 9/17, as labeled in Fig. 3(b). Figures 3(b) and (e) show their appearance in both devices 015 and 002. The $R_{\mathrm{xx}}$ minima, though shallow, appear at the same fractions over a wide range of $D$-field, and are robust in thermal cycling. It is useful to compare their weak but robust appearance with other accidental and irreproducible minima in $R_{xx}$, a number of which can be seen in Figs. 3(c)-(e). Following the same reproducibility criteria, we identify the $\nu = 7/2$ and -1/2 states to be Pfaffian (Fig. 3(c) and (d)). In the vicinity of all four even-denominator states, we find the daughter states of either the Pfaffian or the anti-Pfaffian, but not both, and the daughter states always appear in tandem on both sides of the half-filling. In addition, we see a weak but robust appearance of the $\tilde{\nu} = 7/13$ state near $\nu = -5/2$ (Fig. 11(b)), suggesting that the -5/2 state is likely Pfaffian also. Table 1 summarizes the broken p-h symmetries of five even-denominator states in BLG. Subtle differences of the interaction at different filling factors play a clear role in the resulting asymmetry and calculations capable of explaining all of them self-consistently will shed much theoretical light on this fundamental question. Finally, we briefly note the appearance of unconventional FQH states at $\tilde{\nu} = 2/5$, 3/5, 3/7 and 4/7 in Fig. 3(a), especially the well-developed 2/5 and 3/5 states exhibiting quantized $R_{xy}$ plateaus. These observations offer the future possibility of exploring the proposed non-Abelian orders and topological phase transitions in BLG[4,47].

**C. Valley isospin polarization transitions of odd-denominator fractional quantum Hall states**

The understanding and control of the spin/isospin configuration of an FQH state is a fundamental and key step towards the generation of parafermions[2,7,26,48,49]. The ease to tune $E_v$ and $\nu$ continuously in a single device enables us to systematically study the ground state VIS polarization of FQH states. Figures 4(a) and (b) show maps of $R_{xx}$ ($D$, $\nu$) near $D = 0$ for $4/3 < \nu < 5/3$ and $1/3 < \nu < 2/3$ respectively. In Fig. 4(a), we observe numerous gap closing points reminiscent of spin/pseudospin transitions observed in other 2D systems[2,26,48,49]. Similar gap closing points, and also in the vicinity of $D^*$, were observed in Ref. 12 and interpreted using an effective single-particle-like model[12]. Here we examine our data in the theoretical framework of 2-flux composite fermions (CFs) with SU(2) spin/isospins. In this model, the fractional filling $\nu$ near 3/2 maps to an integer filling of $\nu^*$ filled CF $\Lambda$ levels through $\nu = 2 - \frac{\nu^*}{2\nu^* \pm 1}$. As illustrated in Fig. 4(c), VIS (partial) polarization transitions occur when $\Lambda$ levels of opposite valley indices cross one another. This condition corresponds to $E_v = [1 - \nu^*, 3 - \nu^*, \dots, \nu^* - 1]\hbar\omega_c^*$, for a total number of $\nu^*$ transitions for the $\nu^*$th $\Lambda$ level. Here $\omega_c^* = \frac{eB_{\text{eff}}}{m_{3/2}^*}$ is the cyclotron frequency of CFs, $B_{\text{eff}} = 3(B - B_{3/2})$ and $m_{3/2}^*$ is the effective polarization mass at $\nu = 3/2$[2]. With a single fitting parameter $m_{3/2}^* = 2.6 m_e$, where $m_e$ is the free electron mass in vacuum, we can capture all 8 transitions marked by red circles in Fig. 4(a). In stark contrast to $4/3 < \nu < 5/3$, no valley polarization transitions are observed for states in $1/3 < \nu < 2/3$ in magnetic fields ranging from 14 - 31 T, suggesting all 2-flux CFs in the last LL are spontaneously valley polarized. Following similar argument, we surmise that the even denominator FQH states at $\nu =$ 7/2 and -1/2 may also be spontaneously valley polarized (Fig. 1(c) and Refs. [11,12]) though measurements performed in Fig. 2 for $\nu = 5/2$ are required to confirm this hypothesis. The different behaviors of the FQH states in the two regimes connected by $\nu$ to $2-\nu$ transformation point to strong LL mixing effect[2].

Extending similar measurement and analysis to higher magnetic field (18 - 31 T), we plot in Fig. 4(d) the normalized critical valley Zeeman energy $\alpha^{crit} = E_v/E_c$ corresponding to the onset of full valley polarization (See Fig. 12 in Appendix B for raw data). Here $E_c = \frac{e^2}{4\pi\epsilon\epsilon_0 l_B} = 217\sqrt{B[\mathrm{T}]}$ K is the Coulomb energy scale in BLG using $\epsilon = 3$ for the dielectric constant of h-BN. Results from different magnetic fields collapse quite well, suggesting an approximate $\sqrt{B}$ scaling of $m^*_{3/2}$. The tent-like, solid lines in Fig. 4(d) represent exact diagonalization calculations performed for spin polarization transitions in a zero thickness 2D system[50], which is applicable to any SU(2) isospins. The qualitative agreement between our data and theory supports the SU(2) character of the VIS in BLG. Quantitatively, our results of $\alpha^{crit}$ are more symmetric around 3/2 and are approximately 4 - 5 times smaller than theory. Extrapolation to 3/2 yields $m^*_{3/2} \sim 0.50 m_e\sqrt{B}$, in comparison to the theoretical value of $0.13 m_e\sqrt{B}$ in graphene[50]. Measurements on device 011 yield nearly identical transitions (Fig. 14), indicating that the underlying physics is insensitive to sample details. We attribute the small $\alpha^{crit}$ to the effect of LL mixing, the inclusion of which is necessary to accurately capture the energetics of correlated phenomena in BLG[51].

## IV. CONCLUSION

In summary, we report the observation of an even-denominator FQH state at filling factor 5/2 in Bernal-stacked bilayer graphene. The state remains polarized in the limit of vanishing valley isospin splitting, offering indirect support to a spontaneously spin polarized Moore-Read state. The even-denominator states are particle-hole asymmetric, with the asymmetry consistent with a Pfaffian order at filling factors 3/2 and 7/2, -1/2 and the anti-Pfaffian order at 5/2. The valley isospin behaves like an SU(2) spin with excellent experimental maneuverability. We demonstrate the control of the ground state valley polarization of a large family of FQH states and envision the manipulation of valley isospin to be a powerful tool in elucidating other correlated electronic phenomena and constructing quantum information devices.

## Figure Captions

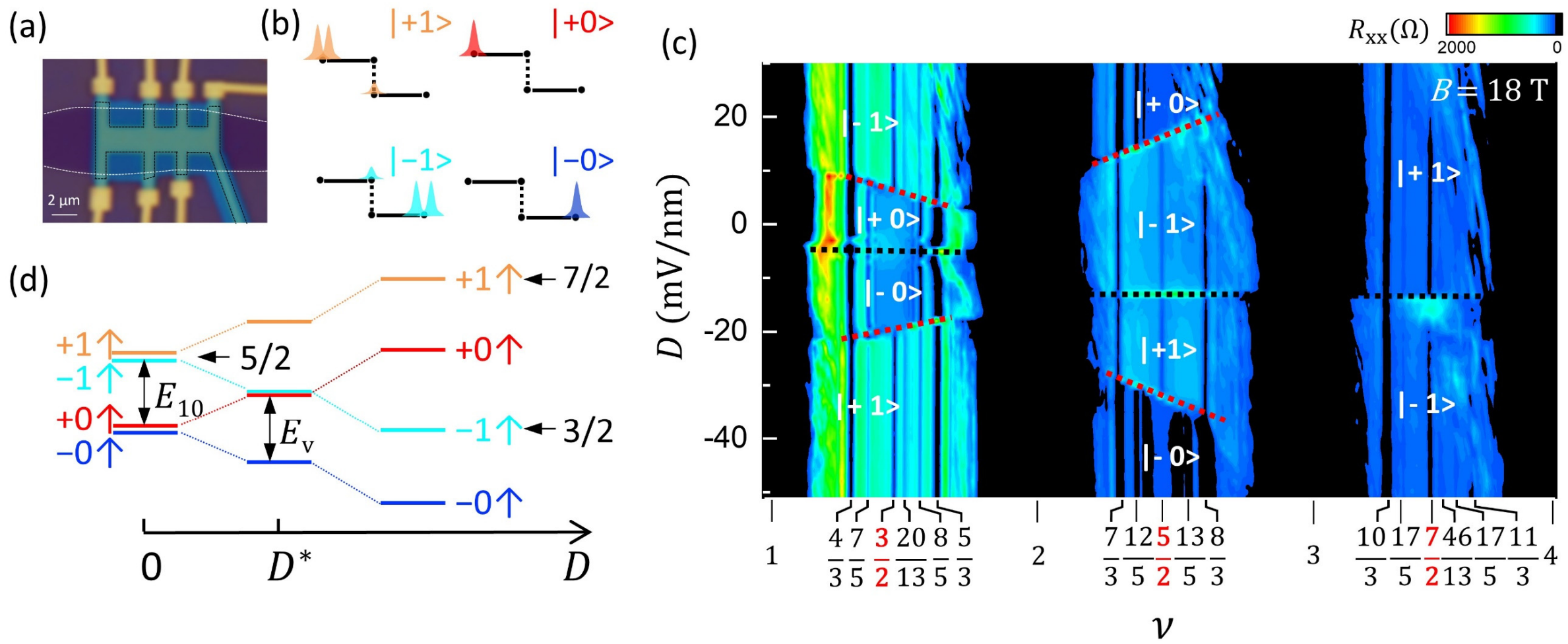

Fig. 1. Valley isospin controlled fractional quantum Hall states in bilayer graphene. (a) An optical micrograph of device 002. The BLG and the top graphite gate are etched into a Hall bar shape outlined in black. The bottom graphite gate is outlined in white. The thickness of the top/bottom BN sheet is 28/23 nm respectively. See Appendix A for fabrication details. (b) illustrates the wave functions of LLs $|\xi, N\rangle$ . $\xi$ = +, - and $N$ = 0, 1 denote the valley and orbital indices. (c) False color map of $R_{xx}$ ($D$, $\nu$) at $B$ = 18 T and $T$ = 20 mK. We label the parent LL of each region. $D^*$ transitions are marked by red dashed lines. The black dashed lines mark the true $D$ = 0 locations. States occupying the $|\pm 0\rangle$ LLs exhibit the 2-flux CF Jain sequence shown in Fig. 10 in Appendix B. (d) An energy level diagram that captures the qualitative features of (c). Even-denominator states at $\nu$=3/2 and 7/2 are reported in Refs. [11-12].

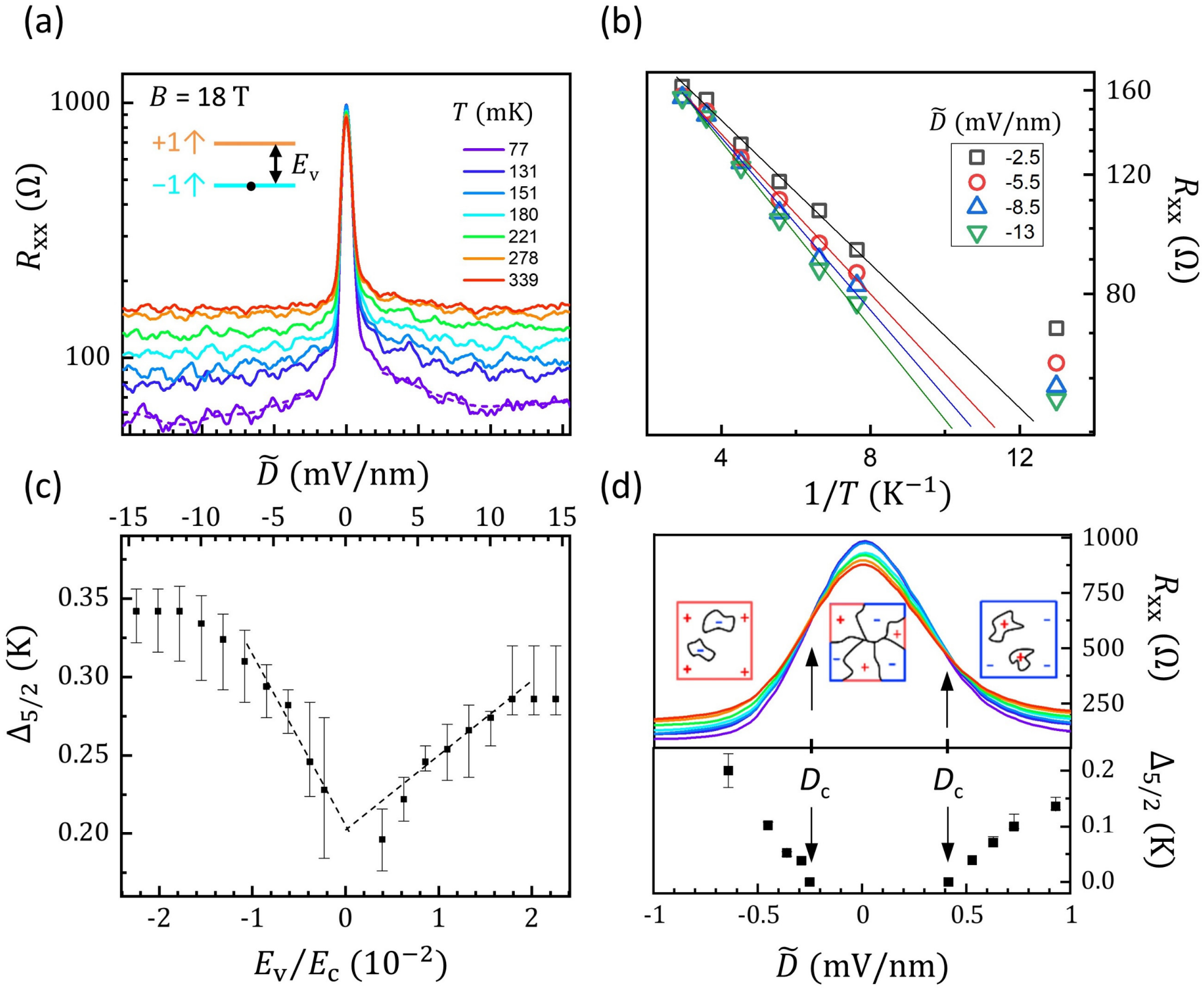

Fig. 2. The gap of the 5/2 state. (a) $R_{xx}$ ($\widetilde{D}$) sweeps at selected temperatures as labeled in the plot. Here both gates are swept simultaneously to stay on the $\nu$ = 5/2 minimum. Data points are read from smoothed traces, an example of which is shown for the 77 mK trace (violet dashed line). (b) $T$-dependent $R_{xx}$ obtained from traces in (a) and plotted on an Arrhenius plot for selected $D$-fields as labeled. Solid lines are fits to exp(-$\Delta_{5/2}/2k_BT$), from which we extract $\Delta_{5/2}$. Upper and lower bounds of $\Delta_{5/2}$from the fits are shown as error bars in (c). Here, we obtain $\widetilde{D} = D + 12.5$ mV/nm using the black dashed line crossing $\nu$ = 5/2 in Fig. 1c. (c) The $\widetilde{D}$-dependence of $\Delta_{5/2}$. Dashed lines extrapolate to 0.2 K at $\widetilde{D}$ = 0. $\Delta_{5/2}$ is larger on the negative $D$ side due to weaker screening from the top graphite gate, which is farther away. (d) Upper panel: The T-dependence of $R_{xx}$ at very small $\widetilde{D}$. As the insets illustrate, the sample is comprised of domains of opposite valley polarizations near $\widetilde{D}$ = 0 and the dominance of bulk conduction leads to $R_{xx} \sim 1/\sigma_{xx}$ and a negative d$R$/d$T$. Transition to edge conduction at large $\widetilde{D}$ produces roughly $T$-independent crossover points at $D_c \sim \pm 0.3$ mV/nm labeled by the arrows. The lower panel of (d) plots the extracted $\Delta_{5/2}$. $\Delta_{5/2}$ drops precipitously towards zero at $D_c$. Similar percolation transitions accompany other valley polarization transitions that occur at both integer and fractional fillings in our samples, such as shown in Fig. 4 and Fig. 12.

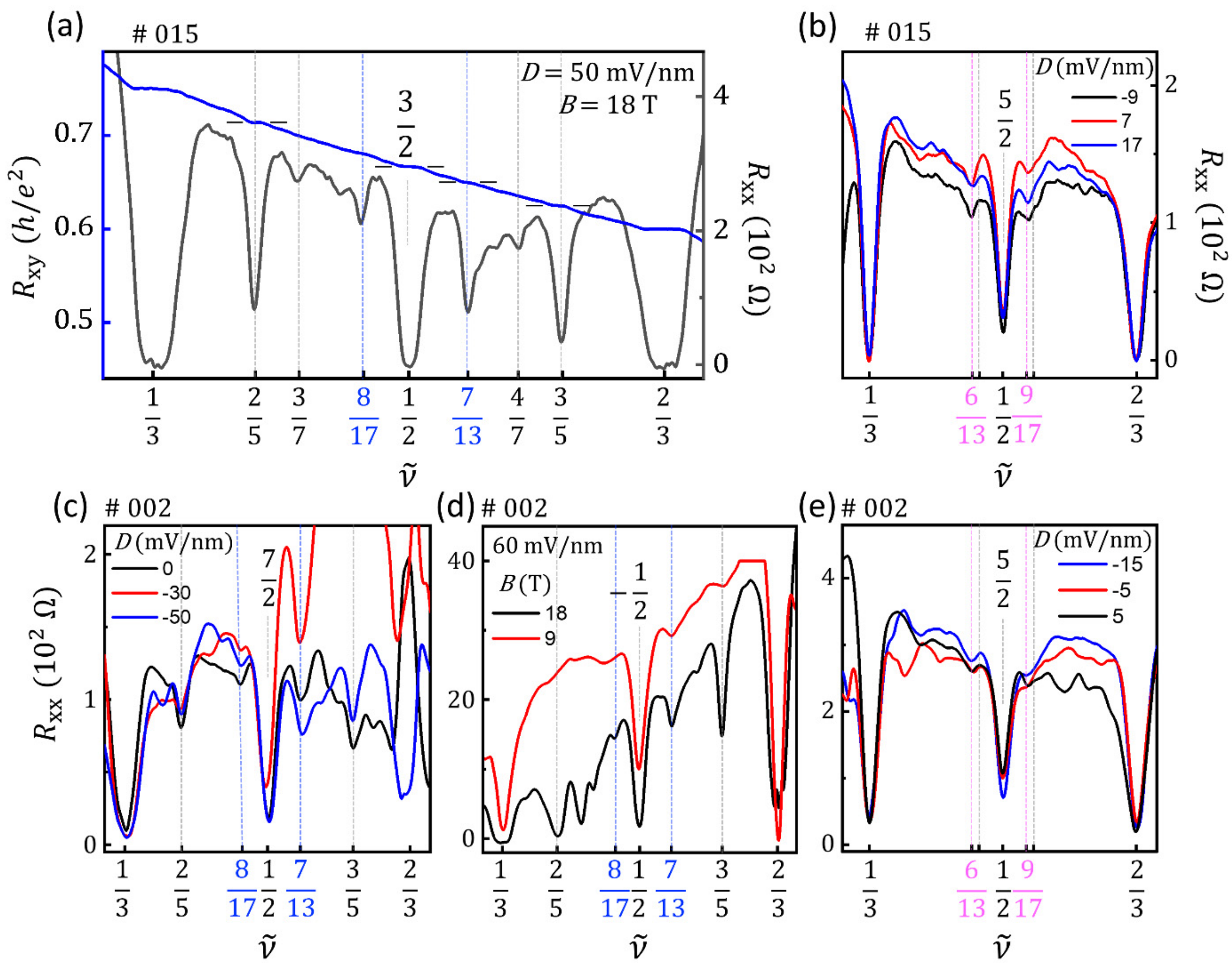

Fig. 3. Particle-hole asymmetry at half-filled N = 1 LLs in BLG. (a) – (e) plot $R_{xx}$ as a function of the partial filling $\tilde{\nu} = \nu$ - [ $\nu$ ] near $\nu$ = 3/2 (a), 5/2 (b), 7/2 (c), -1/2 (d) and 5/2 (e) respectively. (a) also plots $R_{xy}$ measured concomitantly. (a)-(b) are from device 015 and (c)-(e) are from device 002. The data are obtained at fixed D-fields as labeled and $B$ = 18 T unless otherwise noted. $T$ = 20 mK. In (a) $R_{xy}$ plateaus are observed at $\tilde{\nu}$ = 2/5, 1/2, 7/13, and 3/5 and quantized to the correct values given by their full filling factors. Weak but reproducible $R_{xx}$ minima occur at $\tilde{\nu}$ = 8/17, 3/7and 4/7. In (a)-(e), dashed lines mark fractional fillings calculated from the positions of 1/3 and 2/3. Only reproducible minima are marked. Blue lines mark $\tilde{\nu}$ = 8/17 and 7/13 and magenta lines mark $\tilde{\nu}$ = 6/13 and 9/17. Their differences, though small, are discernable in our data. Additional data on $\nu$ = 3/2 and -5/2 from device 002 are given in Fig. 11 of Appendix B.

Table 1: Ground state wave function symmetry of the even-denominator states in BLG

| $\nu$ | 3/2 | 5/2 | 7/2 | -1/2 | -5/2 |
|---|---|---|---|---|---|
| P-H Symmetry | Pf | aPf | Pf | Pf | Pf (likely) |

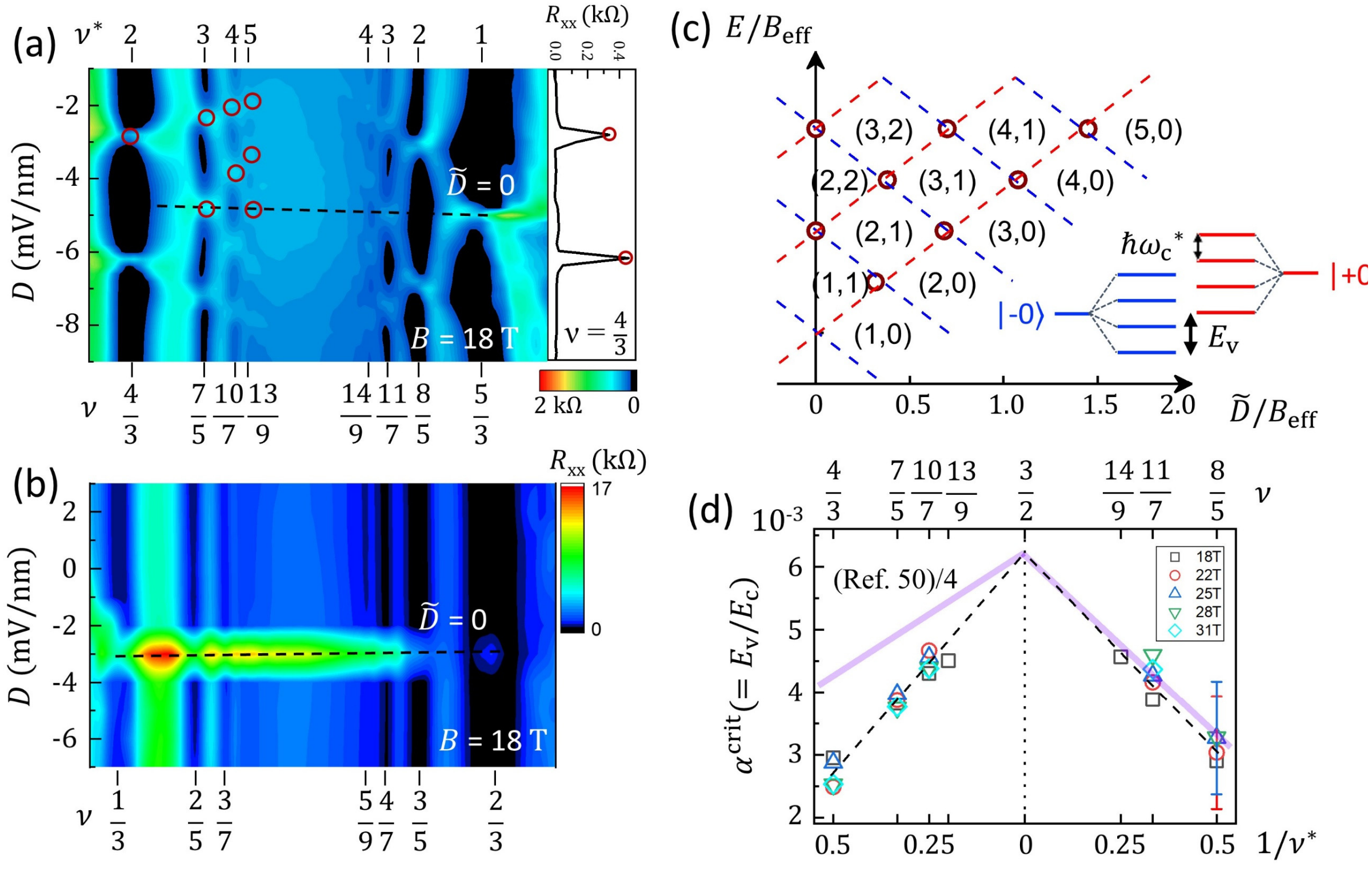

Fig. 4. The valley isospin polarization of Jain FQH states. (a) and (b), False color maps of $R_{xx}$ ($D$, $\nu$) similar to Fig. 1(c) in the filling factor ranges of $4/3 < \nu < 5/3$ (a) and $1/3 < \nu < 2/3$ (b). The map in (a) expands upon the low-D region between the red dashed lines in Fig. 1(c). The top axis in (a) labels the corresponding CF $\Lambda$ level filling factor $\nu^*$. The side panel of (a) plots a line cut through $\nu = 4/3$. The red circles mark $\widetilde{D} \geq 0$ VIS transitions for $\nu$ = 4/3, 7/5, 10/7, and 13/9. Data taken at $B$ = 18 T and $T$ = 20 mK. No transitions are observed in $1/3 < \nu < 2/3$. (c) A free CF $\Lambda$ level fan diagram including valley isospin. The blue and red dash lines represent $\Lambda$ levels of the "+" and "-" valley polarizations respectively and $(n_-, n_+)$ labels the filling factor of each valley. $\nu^* = n_- + n_+$. The slope of the red dashed lines is obtained through a fit to $E_\text{v} = (\nu^* - 1)\hbar\omega_c^*$. This fit yields $m_{3/2}^* = 2.6m_e$, which determines the rest of the diagram. Open circles correspond to transitions marked by the same symbol in (a). Similar analysis is performed for states with $\nu > 3/2$ and yields a smaller $m_{3/2}^* = 1.9m_e$. See Fig. 13 in Appendix B for the fits in both regimes. (d) Normalized critical valley Zeeman energy, $\alpha^\text{crit}$, for FQH states on both sides of $\nu$ = 3/2 at fixed magnetic fields as labeled in the plot. The solid violet lines plot calculations from Ref. [50] divided by 4.

**Acknowledgements:**
K. H., H. F., and J. Z. are supported by the National Science Foundation through grants NSF-DMR-1904986, NSF-DMR-1708972 and by the Kaufman New Initiative research Grant No. KA2018-98553 of the Pittsburgh Foundation. H.F. acknowledges the support of the Penn State Eberly Research Fellowship. D. R. H. and N. A. acknowledge support from the NSF CAREER program (DMR-1654107) and the Penn State 2D Crystal Consortium-Materials Innovation Platform (2DCC-MIP) under NSF Cooperative Agreements DMR-1539916 and DMR-2039351. FIB/SEM and TEM measurements were performed in the Materials Characterization Laboratory of the Materials Research Institute at the Pennsylvania State University. X. L acknowledges the support of Beijing Natural Science Foundation (Grant No. JQ18002) and the National Key Research and Development Program of China (Grant No. 2017YFA0303301). K.W. and T.T. acknowledge support from JSPS KAKENHI (Grant Numbers 19H05790, 20H00354 and 21H05233). Work performed at the National High Magnetic Field Laboratory was supported by the NSF through NSF-DMR-1644779 and the State of Florida. We thank Jainendra Jain, Ajit C. Balram, William Faugno, Zlatko Papic, Bertrand Halperin, Allan H. McDonald, Gabor Csathy for helpful discussions. We thank Dr. Elizabeth Green and Wenkai Zheng for assisting in measurements performed during the COVID-19 pandemic.

**Competing financial interests:**

The authors declare no competing financial interest

## APPEDENX A: DEVICE FABRICARION AND CHARACTERIZATION

### 1. Device fabrication

We fabricated our devices using the processes described below. The top/bottom h-BN sheet thickness is 28 nm/23 nm, 24 nm/27 nm, 35 nm/35 nm, and 25 nm/25nm respectively for devices 002, 011, 012, and 015. We performed transport measurements on devices 002, 011, and 015 and microscopy studies on device 012. Figure 5(a) shows a scanning electron microscopy (SEM) image of device 012. Figure 5(b) and (c) show cross-sectional scanning transmission electron microscopy (STEM) or TEM images of two slabs lifted from the two locations marked in Fig. 5(a) using a focused ion beams (FIB). The sample preparation and microscopy measurements follow that of Ref [32]. TEM studies were performed using an FEI Titan$^3$ G2 operating at 200 kV and an FEI Talos F200X operating at 80 kV. SEM imaging and FIB TEM sample preparation were performed on a Thermo Scientific Scios 2 DualBeam analytical FIB-SEM using ion beam voltages ranging from 30 kV down to 5 kV for lamella thinning.

To make a device we first build a h-BN/graphene/h-BN/bottom graphite gate stack and transfer it to $SiO_2$/Si substrate using a dry transfer method[28,30]. This process is followed by annealing in an Ar/$O_2$ atmosphere at 450 °C for 3 hours to remove polymer residue from the transfer. We then exfoliate a graphite sheet (3 - 4 nm thick) on a PPC (Polypropylene carbonate) stamp and transfer it to the stack. This will be the top gate. Figure 6(a) shows a schematic of the finished stack.

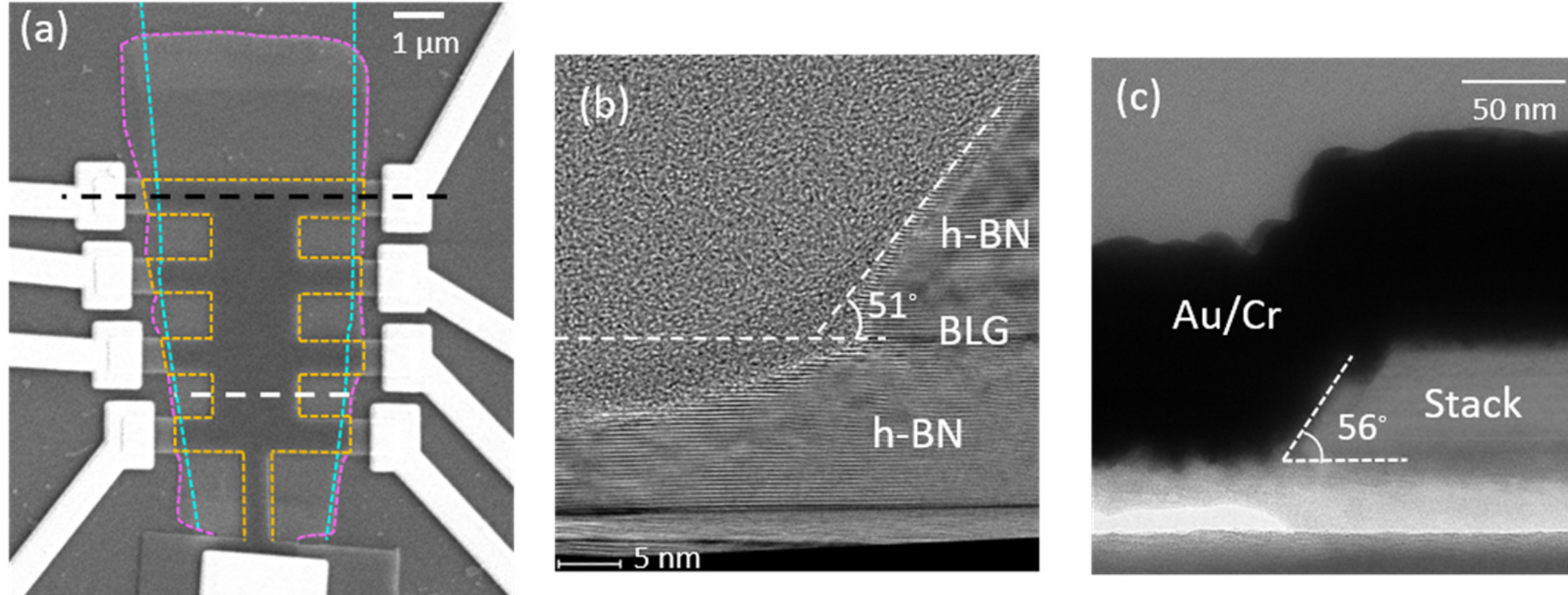

Fig.5. Microscopy studies. (a) An SEM image of device 012. The bottom graphite gate and the bottom h-BN sheet are outlined in cyan and magenta dashed lines respectively. The orange dashed line outlines the Hall bar profile of the top h-BN/BLG stack. (b) High-resolution bright-field STEM image of the cross section cut through the white dashed line in (a). (c) Conventional TEM image of the cross section cut through the black dashed line in (a), showing the Au/Cr side contact. The etching profiles resemble those of previous studies.

A sequence of e-beam lithography and reactive ion etch (RIE) steps illustrated in Figs. 6(b) - (f) is used to shape the stack into the Hall bar structure shown in Fig. 1(a). We first etch the top graphite sheet into an area that is slightly larger than the bottom graphite gate to expose the h-BN/BLG area reserved for making contacts later (Fig. 6(b)). We then use e-beam lithography to define the Hall bar structure (Fig. 6(c)) and three sequential etching steps to pattern the top gate/h-BN/BLG stack. The three steps are illustrated in Figs. 6(d) - (f). The $CHF_3/O_2$ etching step in Fig. 6(d) was done unintentionally for device 002 and repeated in devices 011, 012 and 015. In this step, the top graphite gate protects the h-BN/BLG underneath from being etched and the etching time is not long enough to fully remove the top h-BN sheet in areas outside the top graphite sheet. Device 703 was made without this step and we compare their performances in Fig. 7. We then use $O_2$ plasma to shape the top graphite gate (Fig. 6(e)) followed by a $CHF_3/O_2$ plasma to etch the h-BN/BLG stack into the Hall bar structure (Fig. 6(f)). Table 2 shows the parameters used for the two types of plasma. Etching times for device 002 are shown in Fig. 6 and adjusted for other devices.

| Gas | Pressure (mTorr) | Temperature (°C) | RF Forward Power (W) | Flow (sccm) |
|---|---|---|---|---|
| $O_2$ | 20 | 25 | 14 | 25 |
| $CHF_3$+$O_2$ | 75 | 25 | 40 | $CHF_3$:40<br>$O_2$: 4 |

Table 2: Parameters used in the RIE processes.

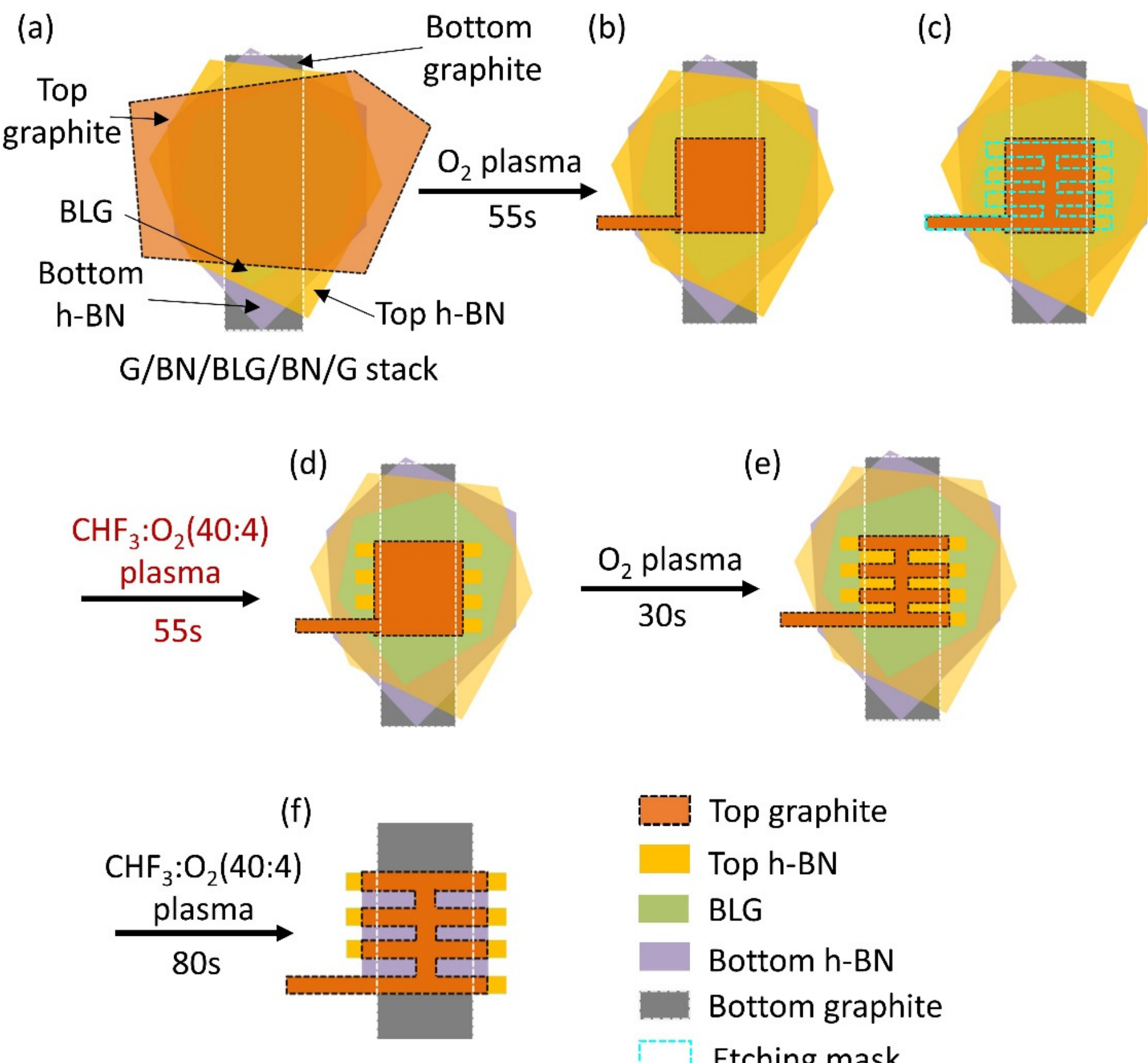

Fig. 6. The etching steps used in the fabrication of device 002. (a) illustrates the finished graphite/h-BN/BLG/h-BN/graphite five-layer stack. (b) illustrates the shape of the top graphite sheet after the $O_2$ etch. It extends over the bottom graphite gate (white dashed lines) by about 200 nm on the left and right sides. (c) overlays the etching mask used in the next three etching steps on (b). (d) The $CHF_3/O_2$ etch defines the shape of the terminals outside the top graphite sheet while the h-BN/BLG underneath the top graphite sheet remains intact. The entire BLG sheet is not exposed in this step. (e) The $O_2$ etch defines the shape of the top graphite gate. (f) The finished Hall bar device.

Finally, we pattern and deposit one-dimensional Au/Cr side contacts using e-beam lithography and physical vapor deposition. The substrate is cooled to 5 ℃ and rotated during the deposition. The deposition rate and thickness for each metal are: Cr: 0.5 Å/s, 5 nm and Au: 1.5 Å/s, 45 nm.

## 2. Device Characterization

Contacts in our devices reside outside the top and bottom graphite gates and are doped by the Si backgate. We apply $V_{Si}$ = 60 V to ensure good performance in a large magnetic field. While STEM images in Figs. 5(b) and (c) did not reveal visible topographic differences compared to previously reported devices[32], electrical contacts in our devices appear to perform better compared to $R_{xx}$ traces shown in the literature, as indicated by less noisy, non-negative values when $R_{xx}$ vanishes at integer and fractional fillings[11,52]. More robust contacts may have facilitated the observations of the even-denominator state at $\nu$ = 5/2, and the valley polarization transitions of the fractional quantum Hall states, which are absent in previous transport studies[11,12]. A second distinguishing feature devices 002 and 011 exhibit is the very narrow valley polarization transition peak at $D$ = 0, which is a strong indicator of higher sample quality. Figure 7(d) compares $R_{xx}$ ($D$) traces taken at the $\nu$ = 2/3 minimum in three devices 002, 011 and 703. Both 002 and 011 are fabricated with the $CHF_3/O_2$ etching step highlighted in red in Fig. 6 while 703 is fabricated without this step. The full-width-at-half-maximum width of the $D$ = 0 peak $\delta D$ is respectively 0.38, 0.24 and 1.43 mV/nm in devices 002, 011 and 703. While it remains unclear to us how the extra etching step reduces $\delta D$, a small disorder broadening of the valley Zeeman splitting is crucial to the observations reported in this work.

We use the top and bottom graphite gates to tune the carrier density $n$ and the perpendicular electric displacement field $D$ following established practices[28,30]. Figure 7(a) plots the longitudinal resistance $R_{xx}$ as a function of the bottom gate voltage $V_{BG}$ at fixed top gate voltage $V_{TG}$'s in device 002. Tracking the charge neutrality points (CNP) ($R_{xx}$ peaks) yields the gating relation $V_{\mathrm{BG}} = -0.81 \times V_{\mathrm{TG}} + 0.012$ as well as ($V_{BG0}$, $V_{TG0}$) corresponding to the $D$ = 0 point marked in Fig. 7(b).

Figure 7(c) plots the Hall resistance $R_{xy}$ vs $n$ at $B$ = 0.05 T in device 002, from which we estimate a disorder-induced density broadening of $\delta n = 7 \times 10^9$ cm$^{-2}$. $\delta n$ is typical of our dual-graphite-gated devices and comparable to what's reported in the literature[11,12]. In Fig. 8, we show that device 002 exhibits a *bulk* disorder energy scale of ~ 10 K, similar to other high quality BLG devices reported in the literature[52]. However, transport in the quantum Hall regime is carried by edge states and as a result sensitively depends upon disorder close to the sample edges. Our devices fabricated using the process illustrated in Fig. 6 show unprecedented quality in this measurement, as indicated by the narrow peaks devices 002 and 011 exhibit at the FQH state $\nu$ = 2/3 in Fig. 7(d).

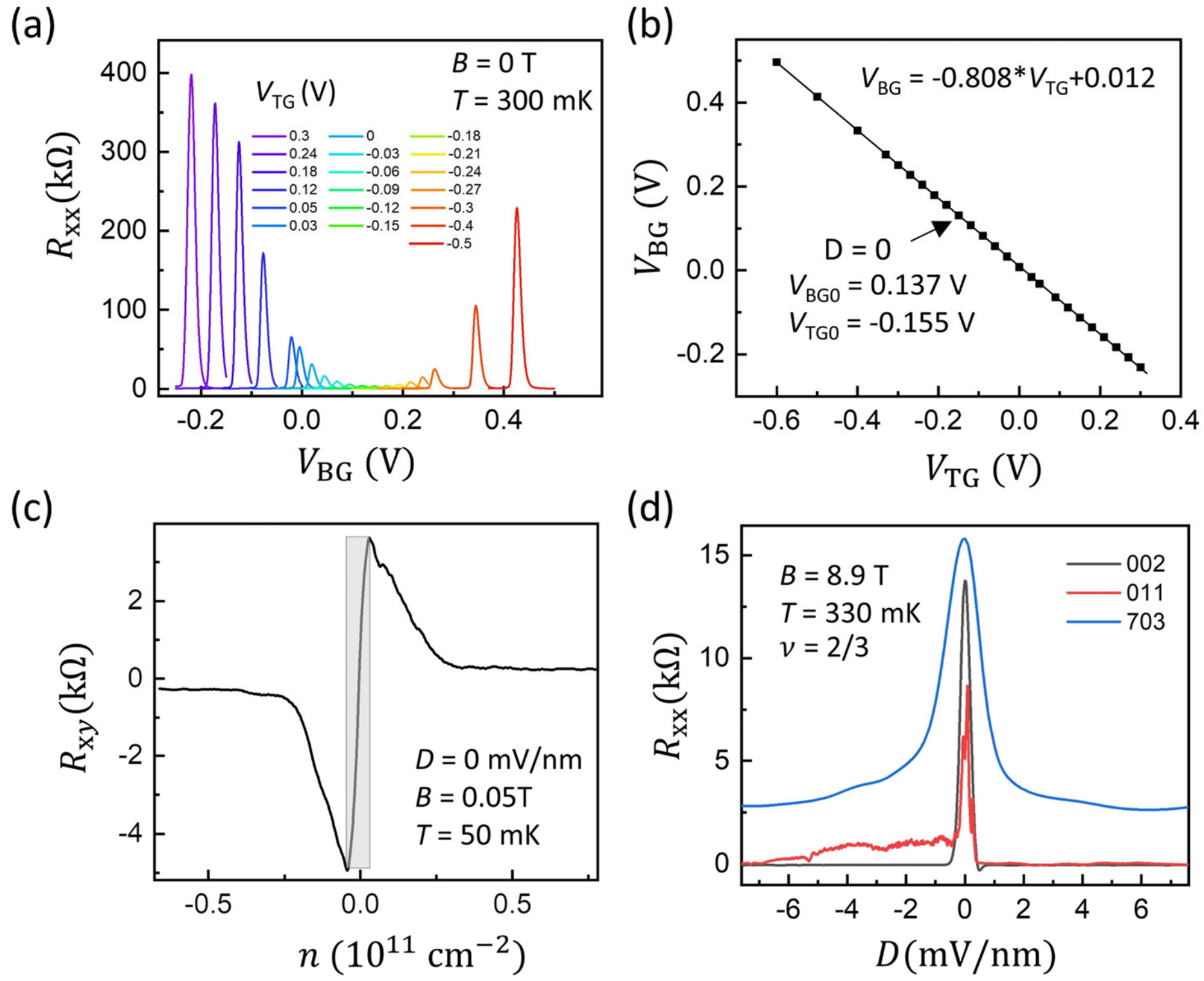

Fig.7. Device characterization. (a) $R_{xx}$ as a function of $V_{BG}$ at fixed $V_{TG}$'s as indicated in the plot. Tracking the peaks allows us to determine the gating relation of $V_{BG}$ and $V_{TG}$, which is plotted in (b). The global minimum occurs at $D = 0$, which corresponds to $V_{BG0} = 0.137$ V and $V_{TG0} = -0.155$ V. (c) The Hall resistance $R_{xy}$ ($n$) at $B = 0.05$ T, $T = 50$ mK and $D = 0$ mV/nm. The shaded region gives an estimated disorder-induced density broadening of $\delta n = 7\times10^9$ cm$^{-2}$. (a) - (c) are from device 002. (d) $R_{xx}$ ($D$) sweeps obtained at the $\nu = 2/3$ minimum in three devices as labeled in the plot. The FWHM width $\delta D = 0.38$, 0.24 and 1.43 mV/nm respectively for devices 002, 011 and 703. Traces are shifted horizontally to overlap the peaks at $D = 0$.

## APPENDIX B: SUPPORTING MEASUREMENTS ON DEVICE 002

### 1. Measurement of the bare valley Zeeman *g*-factor $g_v$

We define the valley Zeeman energy $E_v^{\pm} = \pm 1/2 g_v D$, where ± corresponds to the two valley indices and $g_v$ is the bare valley *g*-factor. The valley Zeeman splitting between the "+" and the "-" levels $E_v = g_v D$. For the $N = 0$ orbital, $E_v = U$, where $U$ is the potential difference between the top and bottom graphene layers. $U$ ($D$) increases linearly with the applied $D$-field and equates the band gap in bilayer graphene $\Delta$ ($D$) at zero magnetic field when $D$ is not too large[53]. We measure $\Delta$ ($D$) at $B = 0$ to determine the bare valley Zeeman *g*-factor $g_v$ using $\Delta(D) = g_v D$. Figures 8(a) - (c) show our measurements and analysis of $\Delta$ ($D$), following approaches described in the supplementary information of Ref. 30 and extending the data of Ref. 30 to the small-$D$ regime. Our analyses yield $g_v = 1.43$ K/mV nm$^{-1}$, in excellent agreement with previous results[30].

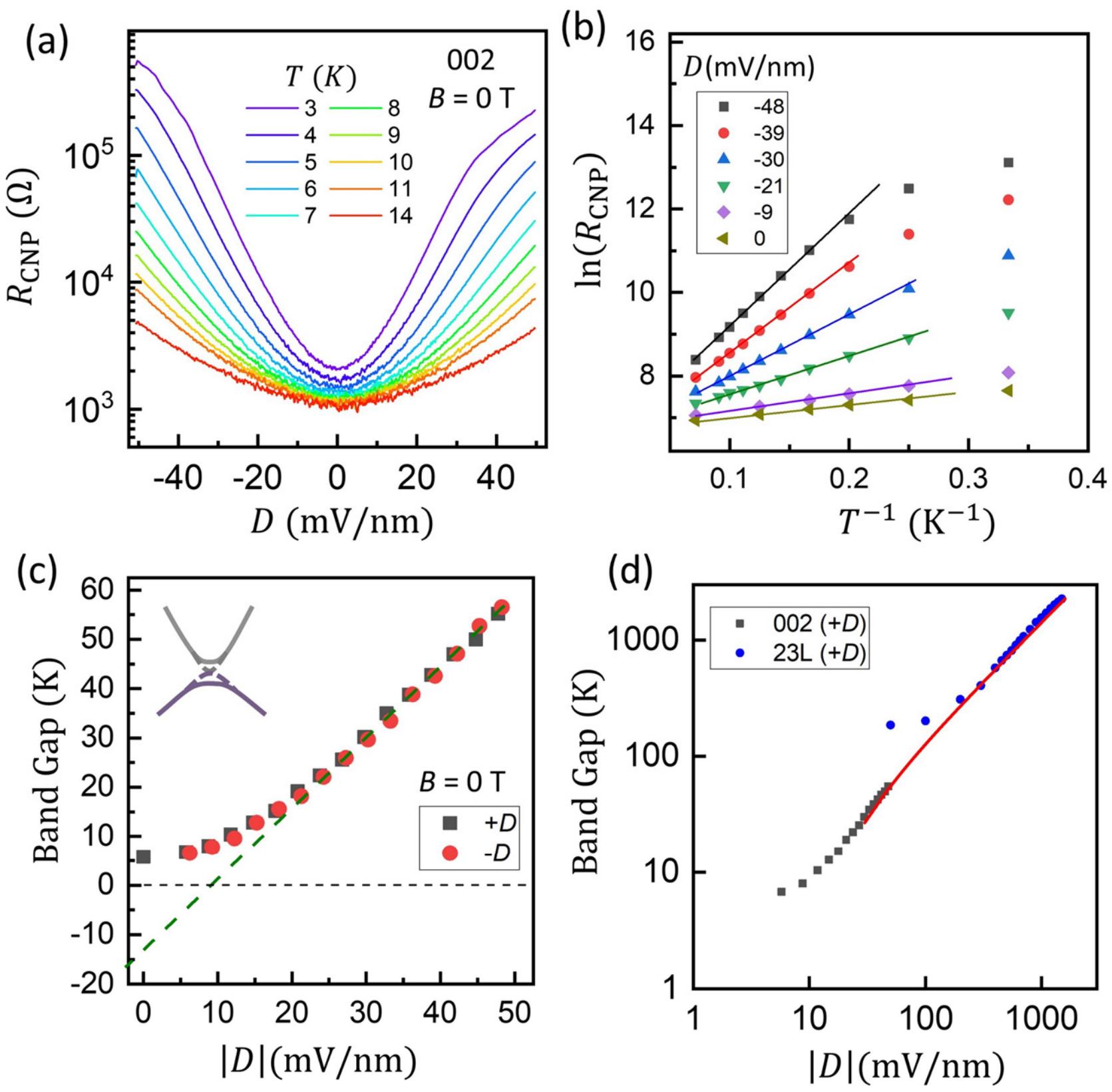

Fig. 8. The $D$-field dependence of the band gap in BLG. (a) The charge neutrality point resistance $R_{CNP}$ as a function of $D$ at different temperatures. (b) $T$-dependent $R_{CNP}$ obtained from traces in (a) and plotted on an Arrhenius plot for selected $D$-fields as labeled. Solid lines are fits to $\exp(-\Delta/2k_BT)$, from which we extract $\Delta$. (c) The extracted $\Delta(D)$ for both positive and negative $D$-fields. We fit data above 20 mV/nm with a line corresponding to $\Delta(D) = 1.43D - 16$ [K] (green dashed line). Disorder leads to the reduction of $\Delta$, which manifests as a negative offset of 16 K. The formation of electron-hole puddles at the CNP gives rise to a finite energy scale of 5 K at $D = 0$. Both point to bulk disorder on the scale of ~ 10 K, similar to other state-of-the-art BLG devices. (d) plots data from devices 002 here and 23 L in Ref. 30. The red line corresponds to the fit obtained in (c).

## 2. FQH states occupying the $N = 0$ and 1 Landau levels

**Figure 9** shows concomitant measurements of $R_{xy}$ and $R_{xx}$ near $\nu = 5/2$ in two devices. Device 015 exhibits a well-quantized plateau in $R_{xy}$ at 18T while device 002 exhibits a developing $R_{xy}$ plateau down to 9T, both confirming the 5/2 state to be a FQH state. **Figure 10** shows examples of FQH states occupying the $N = 0$ LL. They are well described by 2-flux composite fermions. **Figure 11** provides additional evidence to support the breaking of the particle-hole symmetry near even-denominator FQH states $\nu = 3/2$ and -5/2 in bilayer graphene. The Levin-Halperin states manifest as $R_{xx}$ minima at the expected filling factors. They are robust in thermal cycling and appear reproducibly at the same filling factors in different $D$-fields (Fig. 11(a)) and $B$-fields (Fig. 11(b)). We are able to resolve the small differences between the two sets of filling factors, e.g. between 6/13 and 8/17, here and in data shown in Fig. 3 of the main text.

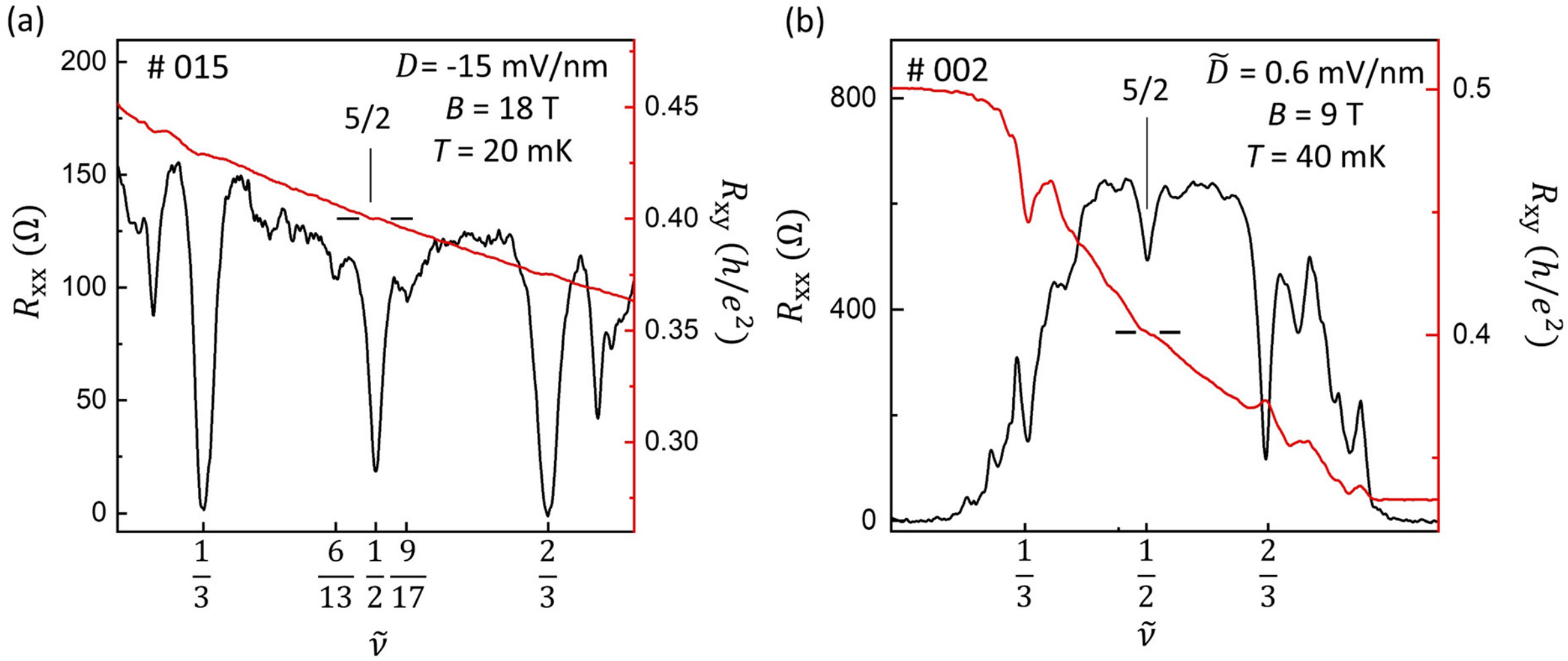

Fig.9. $R_{xx}$ and $R_{xy}$ vs filling factor $\nu$ near 5/2. (a) A deep $R_{xx}$ minimum and a quantized $R_{xy}$ plateau are observed in device 015. (b) In device 002, signatures of the 5/2 state are still present at $B$ = 9T and $T$ = 40 mK.

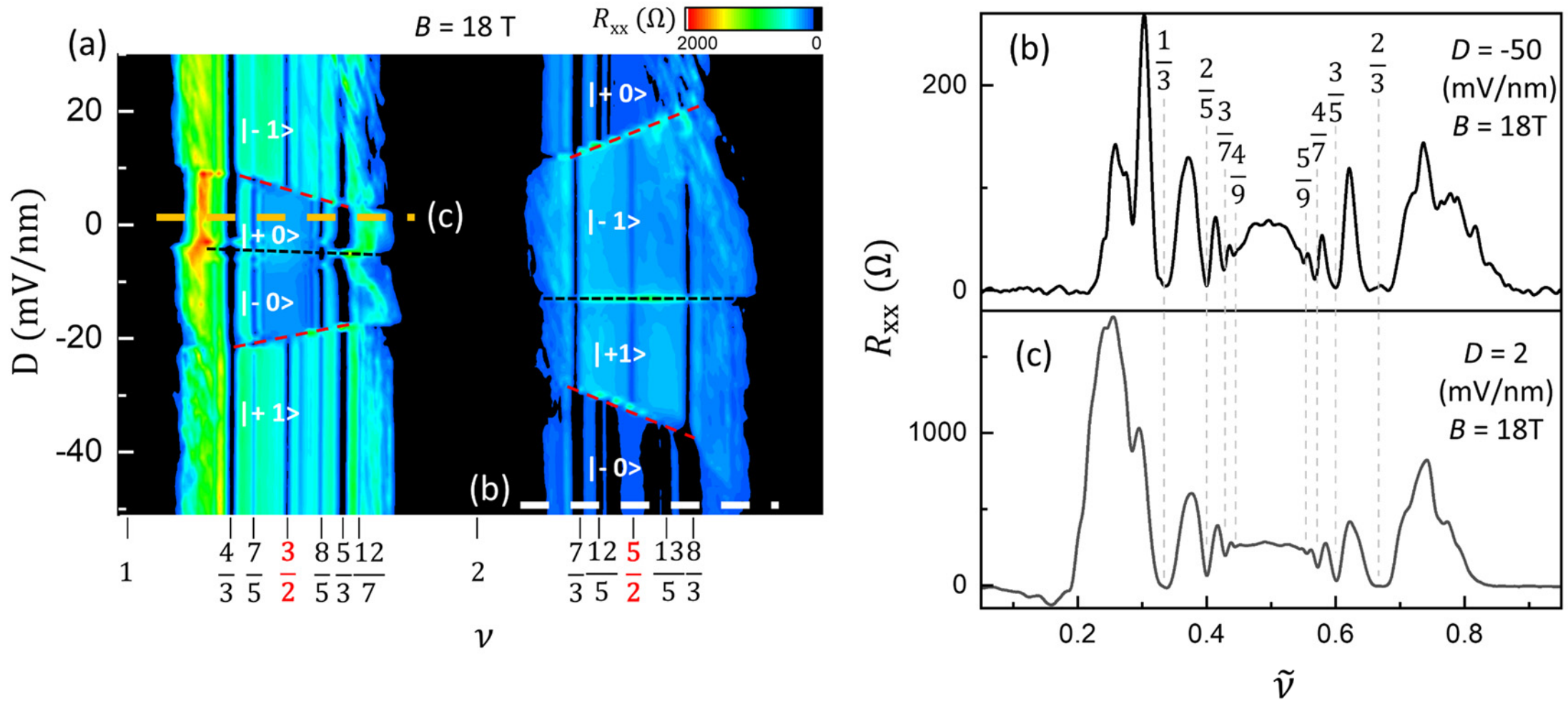

Fig. 10. FQH states on the $N$ = 0 LL. (a) The same false color map of $R_{xx}$ ($D$, $\nu$) as Fig. 1(c) in the main text. (b) shows a trace of $R_{xx}$ ($\nu$) for 2 < $\nu$ < 3 following the white dashed line in (a). (c) shows a trace of $R_{xx}$ ($\nu$) for 1 < $\nu$ < 2 following the orange dashed line in (a). The 2-flux composite fermion FQH states corresponding to $\tilde{\nu} = \nu - [\nu] = \frac{p}{2p\pm1}$, where p is an integer, are labeled in (b) and (c) and provide an excellent description of data. $B$ = 18 T, $T$ = 20 mK.

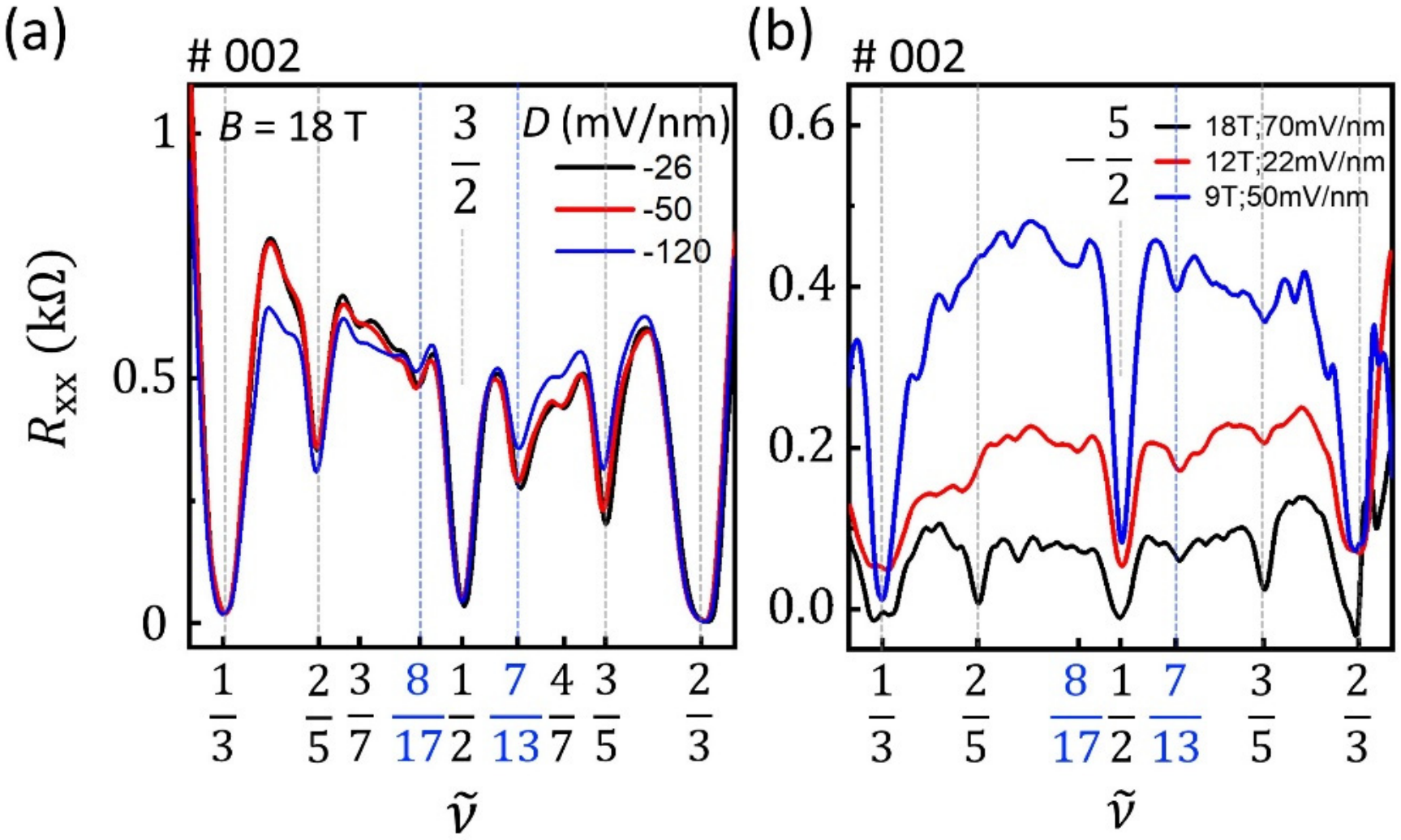

Fig. 11. Additional data from device 002 showing the particle-hole asymmetry near the even-denominator FQH states in bilayer graphene. (a) $R_{xx}$ ($\nu$) at selected $D$-fields in the vicinity of $\nu$ = 3/2 showing the appearance of Levin-Halperin states at $\tilde{\nu}$ = 8/17 and 7/13. (b) $R_{xx}$ ($\nu$) at different $B$- and $D$-fields in the vicinity of $\nu$ = -5/2. The appearance of the weak but reproducible $\tilde{\nu}$ = 7/13 state suggests a Pfaffian state although the $\tilde{\nu}$ = 8/17 is missing. $T$=20 mK.

### 3. Valley isospin polarization transitions of the FQH states near $\nu$ = 3/2

Figure 12 shows the measured $R_{xx}$ ($D$) traces from which we extracted the data points plotted in Fig. 4(d) of the main text. In Fig. 13, we plot transitions to a fully VIS polarized ground state at $B$ = 18 T for filling factors 4/3, 7/5, 10/7 and 13/9 (open red triangles) and 5/3, 8/5, 11/7, 14/9 (open black squares). The data points are read from Fig. 4(a) of the main text and fit to the free CF model $E_{\text{v}}^{\text{crit}} = (\nu^{*} - 1)\,\frac{\hbar e B_{eff}}{m_{3/2}^{*}}$ (Eq. A1), from which we obtained effective polarization mass of $m_{3/2}^{*} = 2.6 m_e$ and $1.9 m_e$ respectively using data from the regimes of $\nu$ < 3/2 and $\nu$ > 3/2.

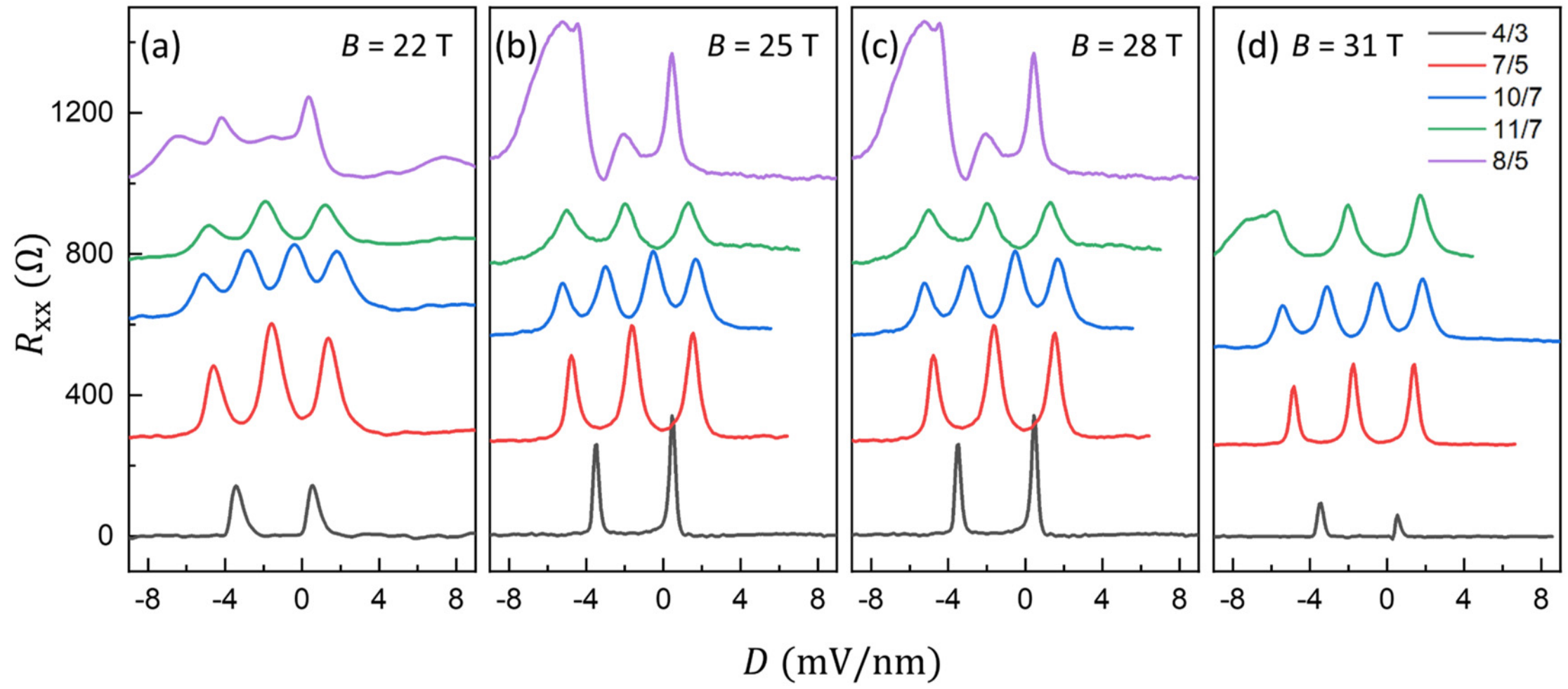

Fig. 12. Valley isospin polarization transitions of the FQH states near $\nu$ = 3/2. (a) - (d) plot traces taken at different $B$-fields as labeled and at different fractions using the color scheme labeled in (d). $T$ = 350 mK. VIS transitions manifest as resistance peaks. The middle peaks of the red trace and the green trace mark the true $D$ = 0 locations. We interpolate/extrapolate them linearly to find the $D$ = 0 locations for the black and the blue traces. Data points in Fig. 4(d) correspond to readings of positive $D$'s from device 002.

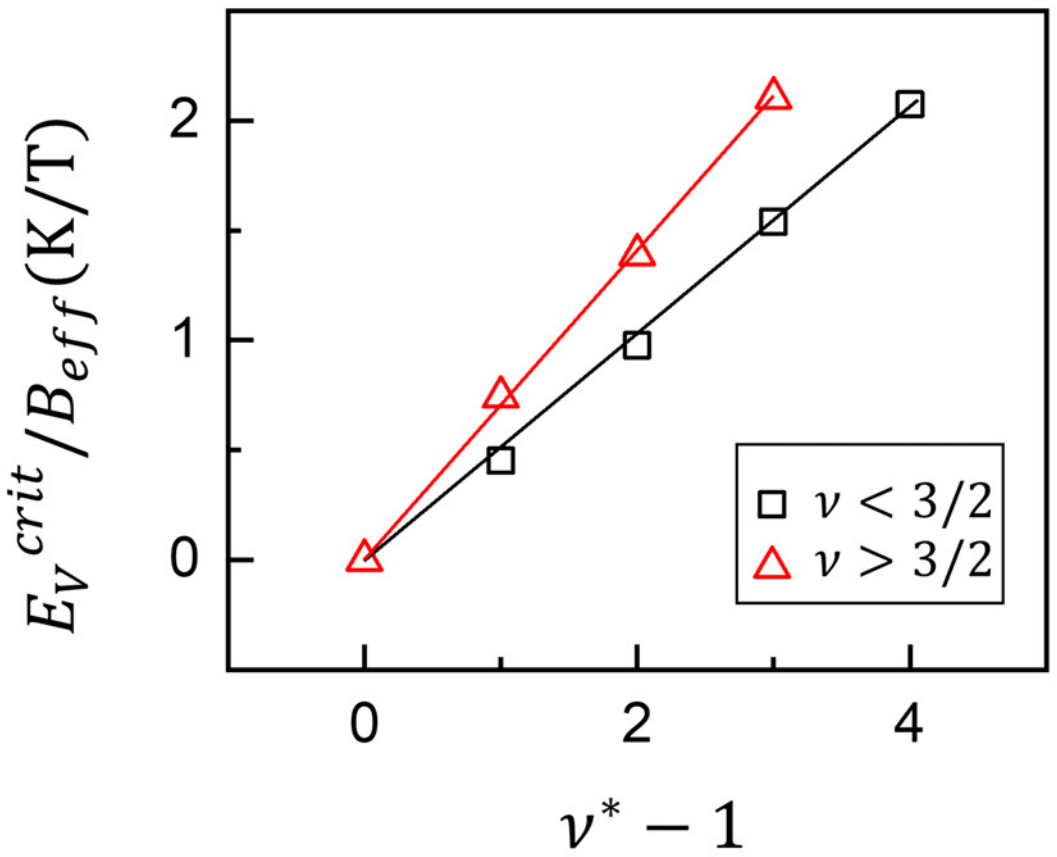

Fig. 13. The effective polarization mass of the CFs at $\nu$ = 3/2. (a) $E_{\mathrm{V}}^{\mathrm{crit}}$ vs $(\nu^* - 1)$ for $\nu$ = 4/3, 7/5, 10/7, 13/9 ($\nu^*$ = 2, 3, 4, 5 respectively, black symbols) and for $\nu$ = 5/3, 8/5, 11/7, 14/9 ($\nu^*$ = 1, 2, 3, 4 respectively, red symbols). The solid lines are fits to Eq. A1. The slopes yield $m_{3/2}^* = 2.6m_e$ for $\nu$ < 3/2 and $m_{3/2}^* = 1.9m_e$ for $\nu$ > 3/2.

## APPENDIX C: DATA FROM DEVICE 011

This section presents measurements from device 011. Device 011 was fabricated using the processes described in appendix A and has top and bottom h-BN sheets with similar thickness to that of device 002. Figure 14 compares the valley isospin polarization transitions in devices 002 and 011 at two fractional states. The transitions occur at nearly identical *D*-fields, indicating the underlying physics is governed by the intrinsic LL structure of bilayer graphene and is insensitive to sample-specific details

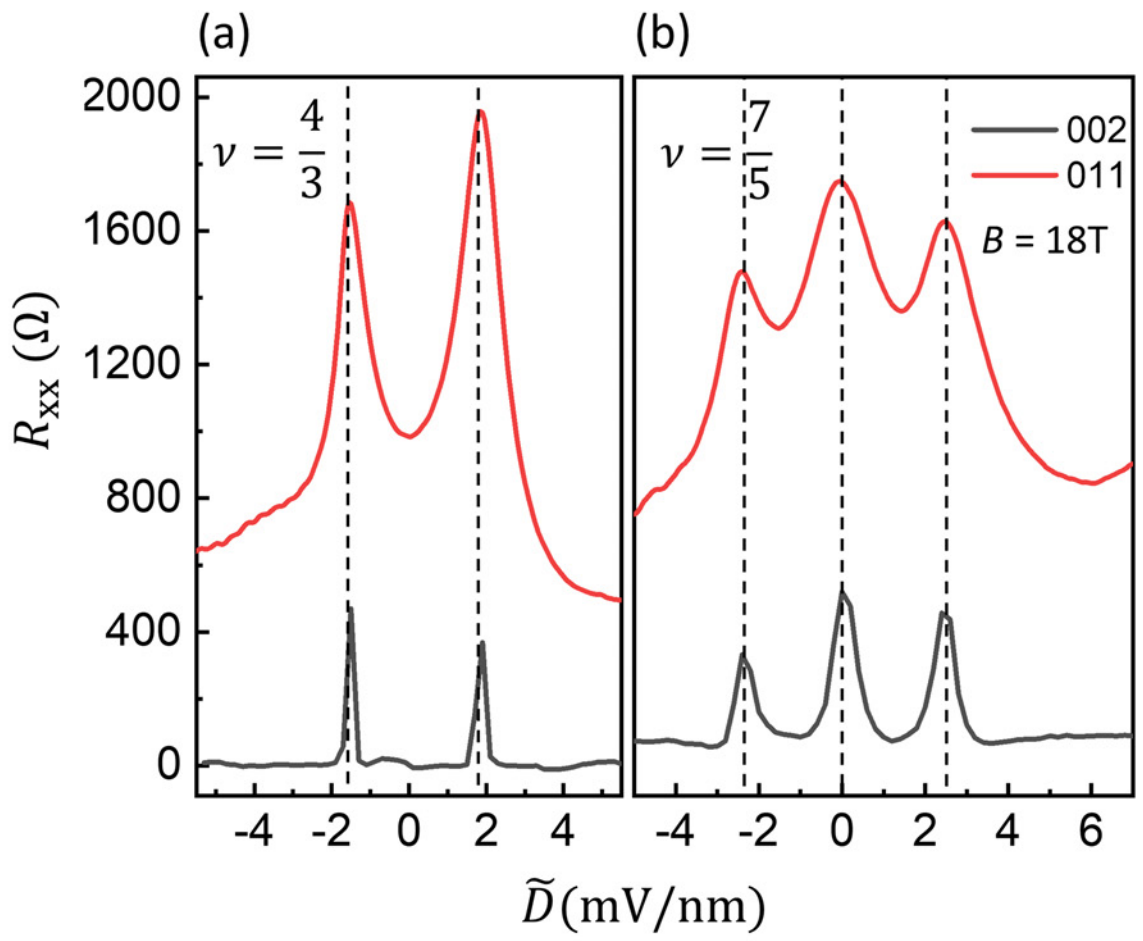

Fig. 14. Comparison of VIS polarization transitions in devices 002 (black traces) and 011 (red traces) at $\nu$ = 4/3 (panel (a)) and 7/5 (panel (b)). $B$ = 18T and $T$ = 20 mK. Device 011 exhibits very narrow VIS transitions shortly after the device was fabricated (See Fig. 7(d)). Its quality degraded over many months of storage before measurements shown here were taken. Traces plotted here are 3-terminal measurements so the contact resistance is included. Nonetheless, nearly identical VIS polarization transitions are found in these two devices, indicating the underlying physics is insensitive to sample details and relatively robust against disorder.

## References

1 Das Sarma, S. & Pinczuk, A. Eds. *Perspectives in Quantum Hall Effects: Novel Quantum Liquids in Low-Dimensional Semiconductor Structures*. (Wiley, 1997).

2 Halperin, B. & Jain, J. K. Eds. *Fractional Quantum Hall Effects: New Developments* (World Scientific, 2020).

3 Moore, G. & Read, N. *Nonabelions in the fractional quantum hall effect*. Nucl. Phys. B **360**, 362-396 (1991).

4 Read, N. & Rezayi, E. *Beyond paired quantum Hall states: Parafermions and incompressible states in the first excited Landau level*. Phys. Rev. B **59**, 8084-8092 (1999).

5 Nayak, C., Simon, S. H., Stern, A., Freedman, M. & Das Sarma, S. *Non-Abelian anyons and topological quantum computation*. Rev. Mod. Phys. **80**, 1083-1159 (2008).

6 Willett, R. L. *The quantum Hall effect at 5/2 filling factor*. Rep. Prog. Phys. **76**, 076501 (2013).

7 Mong, R. S. K. *et al. Universal Topological Quantum Computation from a Superconductor-Abelian Quantum Hall Heterostructure*. Phys. Rev. X **4**, 011036 (2014).

8 Nakamura, J., Liang, S., Gardner, G. C. & Manfra, M. J. *Direct observation of anyonic braiding statistics*. Nat. Phys. **16**, 931-936 (2020).

9 Bartolomei, H. *et al. Fractional statistics in anyon collisions*. Science **368**, 173 (2020).

10 Morf, R. H. *Transition from quantum Hall to compressible states in the second Landau level: New light on the* ν*=5/2 enigma*. Phys. Rev. Lett. **80**, 1505-1508 (1998).

11 Li, J. I. A. *et al. Even-denominator fractional quantum Hall states in bilayer graphene*. Science **358**, 648-651 (2017).

12 Zibrov, A. A. *et al. Tunable interacting composite fermion phases in a half-filled bilayer-graphene Landau level*. Nature **549**, 360-364 (2017).

13 Ki, D. K., Fal'ko, V. I., Abanin, D. A. & Morpurgo, A. F. *Observation of Even Denominator Fractional Quantum Hall Effect in Suspended Bilayer Graphene*. Nano Lett. **14**, 2135-2139 (2014).

14 Falson, J. *et al. Even-denominator fractional quantum Hall physics in ZnO*. Nat. Phys. **11**, 347-351 (2015).

15 Shi, Q. *et al. Odd- and even-denominator fractional quantum Hall states in monolayer WSe2*. Nat. Nano. **15**, 569-573 (2020).

16 Banerjee, M. *et al. Observation of half-integer thermal Hall conductance*. Nature **559**, 205-210 (2018).

17 Lee, S.-S., Ryu, S., Nayak, C. & Fisher, M. P. A. *Particle-Hole Symmetry and the* ν *= 5/2 Quantum Hall State*. Phys. Rev. Lett. **99**, 236807 (2007).

18 Levin, M., Halperin, B. I. & Rosenow, B. *Particle-Hole Symmetry and the Pfaffian State*. Phys. Rev. Lett. **99**, 236806 (2007).

19 Son, D. T. *Is the Composite Fermion a Dirac Particle?* Phys. Rev. X **5**, 031027 (2015).

20 Ma, K. K. W. & Feldman, D. E. *The sixteenfold way and the quantum Hall effect at half-integer filling factors*. Phys. Rev. B **100**, 035302 (2019).

21 Kumar, A., Csathy, G. A., Manfra, M. J., Pfeiffer, L. N. & West, K. W. *Nonconventional Odd-Denominator Fractional Quantum Hall States in the Second Landau Level*. Phys. Rev. Lett. **105**, 246808 (2010).

22 Dutta, B. *et al. Distinguishing between non-abelian topological orders in a Quantum Hall system*. Science **375**, 193–197 (2022).

23 Dimov, I., Halperin, B. I. & Nayak, C. *Spin order in paired quantum Hall states*. Phys. Rev. Lett. **100**, 126804 (2008).

24 Feiguin, A. E., Rezayi, E., Yang, K., Nayak, C. & Das Sarma, S. *Spin polarization of the* ν *= 5/2 quantum Hall state*. Phys. Rev. B **79**, 115322 (2009).

25 Eisenstein, J. P., Boebinger, G. S., Pfeiffer, L. N., West, K. W. & He, S. *New fractional quantum Hall state in double-layer two-dimensional electron systems*. Phys. Rev. Lett. **68**, 1383-1386 (1992).

26 Padmanabhan, M., Gokmen, T. & Shayegan, M. *Density dependence of valley polarization energy for composite fermions*. Phys. Rev. B **80**, 035423 (2009).

27 Li, J. I. A., Taniguchi, T., Watanabe, K., Hone, J. & Dean, C. R. *Excitonic superfluid phase in double bilayer graphene*. Nat. Phys. **13**, 751-755 (2017).

28 Li, J. *et al. A valley valve and electron beam splitter*. *Science* **362**, 1149 (2018).

29 Hunt, B. M. *et al. Direct measurement of discrete valley and orbital quantum numbers in bilayer graphene*. Nat. Commun. **8**, 948 (2017).

30 Li, J., Tupikov, Y., Watanabe, K., Taniguchi, T. & Zhu, J. *Effective Landau Level Diagram of Bilayer Graphene*. Phys. Rev. Lett. **120**, 047701 (2018).

31 Li, J. *et al. Metallic Phase and Temperature Dependence of the* ν *= 0 Quantum Hall State in Bilayer Graphene*. Phys. Rev. Lett. **122**, 097701 (2019).

32 Wang, L. *et al. One-Dimensional Electrical Contact to a Two-Dimensional Material*. Science **342**, 614 (2013).

33 Eisenstein, J. P. *et al. Collapse of the Even-Denominator Fractional Quantum Hall Effect in Tilted Fields*. Phys. Rev. Lett. **61**, 997-1000 (1988).

34 Tiemann, L., Gamez, G., Kumada, N. & Muraki, K. *Unraveling the Spin Polarization of the ν=5/2 Fractional Quantum Hall State*. Science **335**, 828-831 (2012).

35 Liu, G. T. *et al. Enhancement of the* ν *=5/2 Fractional Quantum Hall State in a Small In-Plane Magnetic Field*. Phys. Rev. Lett. **108**, 196805 (2012).

36 Pan, W. *et al. Competing quantum Hall phases in the second Landau level in the low-density limit*. Phys. Rev. B **89**, 241302 (2014).

37 Samkharadze, N., Ro, D., Pfeiffer, L. N., West, K. W. & Csáthy, G. A. *Observation of an anomalous density-dependent energy gap of the* ν *=5/2 fractional quantum Hall state in the low-density regime*. Phys. Rev. B **96**, 085105 (2017).

38 Yoo, H. M., Baldwin, K. W., West, K., Pfeiffer, L. & Ashoori, R. C. *Complete spin phase diagram of the fractional quantum Hall liquid*. Nat. Phys. **16**, 1022–1027 (2019).

39 Wang, P. *et al. Finite-thickness effect and spin polarization of the even-denominator fractional quantum Hall states*. Phys. Rev. Res. **2**, 022056 (2020).

40 Dean, C. R. *et al. Contrasting Behavior of the 5/2 and 7/3 Fractional Quantum Hall Effect in a Tilted Field*. Phys. Rev. Lett. **101** (2008).

41 Zaletel, M. P., Mong, R. S. K., Pollmann, F. & Rezayi, E. H. *Infinite density matrix renormalization group for multicomponent quantum Hall systems*. Phys. Rev. B **91**, 045115 (2015).

42 Pakrouski, K. *et al. Phase Diagram of the* ν *= 5/2 Fractional Quantum Hall Effect: Effects of Landau-Level Mixing and Nonzero Width*. Phys. Rev. X **5**, 021004 (2015).

43 Rezayi, E. H. *Landau Level Mixing and the Ground State of the* ν *= 5/2 Quantum Hall Effect*. Phys. Rev. Lett. **119**, 026801 (2017).

44 Zucker, P. T. & Feldman, D. E. *Stabilization of the Particle-Hole Pfaffian Order by Landau-Level Mixing and Impurities That Break Particle-Hole Symmetry*. Phys. Rev. Lett. **117**, 096802 (2016).

45 Zhu, W. & Sheng, D. N. *Disorder-Driven Transition in the* ν *=5/2 Fractional Quantum Hall Effect*. Phys. Rev. Lett. **123**, 056804 (2019).

46 Levin, M. & Halperin, B. I. *Collective states of non-Abelian quasiparticles in a magnetic field*. Phys. Rev. B **79**, 205301 (2009).

47 Faugno, W. N., Jain, J. K. & Balram, A. C. *Non-Abelian fractional quantum Hall state at 3/7-filled Landau level*. Phys. Rev. Res. **2**, 033223 (2020).

48 Du, R. R. *et al. Fractional Quantum Hall Effect around* ν *= 3/2: Composite Fermions with a Spin*. Phys. Rev. Lett. **75**, 3926-3929 (1995).

49 Feldman, B. E. *et al. Fractional Quantum Hall Phase Transitions and Four-Flux States in Graphene*. Phys. Rev. Lett. **111**, 076802 (2013).

50 Balram, A. C., Tőke, C., Wójs, A. & Jain, J. K. *Fractional quantum Hall effect in graphene: Quantitative comparison between theory and experiment*. Phys. Rev. B **92**, 075410 (2015).

51 Papić, Z. & Abanin, D. A. *Topological Phases in the Zeroth Landau Level of Bilayer Graphene*. Phys. Rev. Lett. **112**, 046602 (2014).

52 Rhodes, D., Chae, S. H., Ribeiro-Palau, R. & Hone, J. *Disorder in van der Waals heterostructures of 2D materials*. *Nat. Mater.* **18**, 541-549 (2019).

53 McCann, E. & Koshino, M. *The electronic properties of bilayer graphene*. Rep. Prog. Phys. **76**, 056503 (2013).